# Polarization switching on the open surfaces of the wurtzite ferroelectric nitrides: ferroelectric subsystems and electrochemical reactivity

Yongtao Liu,[1*] Anton V. Ievlev,[1] Eugene A. Eliseev,[2] Nana Sun,[3] Kazuki Okamoto,[3] Hiroshi Funakubo,[3] Anna N. Morozovska,[4] Sergei Kalinin[5#]

[1] Center for Nanophase Materials Sciences, Oak Ridge National Laboratory, Oak Ridge, TN 37830, USA
[2] Frantsevich Institute for Problems in Materials Science, National Academy of Sciences of Ukraine, Omeliana Pritsaka str., 3, Kyiv, 03142 Ukraine
[3] Department of Material Science and Engineering, School of Materials and Chemical Technology, Institute of Science Tokyo, Yokohama, 226-8502, Japan
[4] Institute of Physics, National Academy of Sciences of Ukraine, 46, pr. Nauky, Kyiv, 03028 Ukraine
[5] Department of Materials Science and Engineering, University of Tennessee, Knoxville, TN 37996, USA.

* liuy3@ornl.gov; [#] sergei2@utk.edu

**Abstract**

Binary ferroelectric nitrides are promising materials for information technologies and power electronics. However, polarization switching in these materials is highly unusual. From the structural perspective, polarization reversal is associated with the change of the effective polarity at the surfaces and interfaces from N-to-M terminated, suggesting strong coupling between ferroelectric and chemical phenomena. Phenomenologically, macroscopic studies demonstrate the presence of complex time dependent phenomena including wake-up. Here, we explore the polarization switching using the multidimensional high-resolution piezoresponse force microscopy (PFM) and spectroscopy, detecting both the evolution of induced ferroelectric domain, electromechanical response, and surface deformation during first-order reversal curve measurements. We demonstrate the presence of two weakly coupled ferroelectric subsystems and the bias-induced electrochemical reactivity. The observed behaviors are very similar to the recent studies of other wurtzite system but additionally include electrochemical reactivity, suggesting the universality of these behaviors for the wurtzite binary ferroelectrics. These studies suggest potential of high-resolution multimodal PFM spectroscopies to resolve complex coupled polarization dynamics in materials. Furthermore, these PFM based studies are fully consistent with the recent electron microscopy observations of the shark-teeth like ferroelectric domains in nitrides. Hence, we believe that these studies establish the universal phenomenological picture of polarization switching in binary wurtzite.

The discovery of ferroelectricity in binary fluorites such as hafnia ($HfO_2$),[1] and more recently, in wurtzite-structured materials such as $Al_{1-x}Sc_xN$,[2] $Al_{1-x}B_xN$,[3] and $Zn_{1-x}Mg_xO$,[4] has transformed the landscape of ferroelectric research. Hafnia and wurtzite-based ferroelectrics are compatible with standard complementary metal-oxide-semiconductor processes, offering a clear path toward scalable integration in microelectronic devices. Despite rapid growth of research effort in this area, the high coercive fields of wurtzite ferroelectrics remain the major limiting factor for the practical deployment of these materials in memory technologies, highlighting the need for a deeper exploration of the switching mechanisms at multiple length scales to uncover its origin.

From the first studies of the ferroelectricity in these binary systems more than decade ago,[2] it has been realized that polarization switching in these systems is accompanied by unique and often perplexing physical phenomena in macroscopic ferroelectric characterization. These include wake up effects,[5] when ferroelectric responses emerge only after significant number of cycles, highly unusual nucleation mechanisms as revealed by switching kinetics analysis,[6-8] etc. From the mechanistic viewpoint, the mechanisms of polarization switching and domain formation in binary ferroelectrics diverge significantly from those in classical perovskite ferroelectrics.[9-12] In classical perovskite ferroelectrics, polarization switching is associated with the movement of the central cation within the oxygen octahedra, white the chemical nature of surfaces and interfaces remains the same. While polarization couples to reactivity in processes such as acid dissolution or photoelectrochemistry,[13-15] surface terminations can remain stable. Conversely, in wurtzite nitrides and oxides, switching by definition is associated with changes in surface or interface termination from anionic to cationic species, implying substantial reconfiguration of local chemical bonding during the switching process. This indicates a strong interplay between polarization, lattice structure, and electrochemical processes, particularly at boundaries and defects. Similarly, the presence of topological defects may suggest the possibility of complex and perhaps non-traditional domain dynamics, but direct experimental evidence remains limited.

Recent advances in high-resolution characterization techniques, especially scanning transmission electron microscopy (STEM),[16-18] have begun to shed light on these mesoscale mechanisms. Notably, recent observations of so-called "dragon teeth" domain structures in wurtzite ferroelectrics reveal a highly nontrivial domain topology,[18] possibly indicative of localized structural phase transitions and the coexistence of multiple polarization states. These findings underscore the importance of bridging the gap between atomic-scale mechanisms and macroscopic polarization switching, particularly in materials where classical ferroelectric models do not readily apply. In our previous studies on the $Zn_{1-x}Mg_xO$ system, we identified evidence for a fringing ridge polarization switching process,[8] suggesting a more intricate internal landscape than previously appreciated.

Building upon this foundation, in this work, we investigate local polarization switching behavior in wurtzite nitrides, with a particular focus on their mesoscale mechanisms. Our results reveal similar switching behavior, these results support the hypothesis that the fringing ridge domain switching mechanism is a universal feature of wurtzite ferroelectrics. However, we also revealed a notable distinction in wurtzite nitrides, where switching is accompanied by clear signatures of electrochemical activity.[19] The complementary data from the electromechanical response and surface displacement measurements allows to separate these coupled mechanisms.

This observation suggests that electrochemical processes may be fundamentally coupled to polarization dynamics in these materials, particularly under high electric fields.

We investigate a 40 nm thick $Al_{0.8}Sc_{0.2}N$ thin film. The $Al_{0.8}Sc_{0.2}N$ film was prepared by the dual-source radio frequency (RF) reactive magnetron sputtering method from Al and Sc metal targets on (111)Pt/Ti/SiO$_2$/(001)Si substrates.[20] Details of the film deposition were reported elsewhere.[20, 21] To get the insight into the polarization switching, we first investigated classical polarization switching waveforms using Band Excitation Piezoresponse Spectroscopy (BEPS),[22] as illustrated in Figure 1. BEPS measurement was performed across a 5×5 location array, the BEPS waveform (Figure 1a) comprises 10 cycles of a triangular excitation waveform, this waveform included both on-field and off-field conditions at each step and had a maximum magnitude of ±40 V. The averaged BEPS loop obtained from this measurement is shown in Figure 1b. At the onset of the BEPS loop (indicated in cyan), the loops exhibit an enhancement of piezoresponse as the applied DC voltage sweeps from 0 to +40 V, followed by a decrease in piezoresponse when the voltage sweeps to -40 V. In subsequent cycles, the loops demonstrate clear polarization switching, with the piezoresponse stabilizing and remaining nearly constant. The topography measured after the BEPS experiments (Figure 1c) reveals protruding structures in the regions where BEPS was conducted.

The evolution of the sample topography during BEPS measurements was tracked using the height channel, as indicated by the black curve in Figure 1a. Note that topographic tracing is traditionally ignored in PFM spectroscopy measurements; however the capability of scanning probe microscopy to detect topographic changes on the sub-nm level potentially allows to detect changes in material volume due to Vegard expansion, changes in surface terminations, or second phase formation. These strain measurements are complementary to dynamic electromechanical measurements since they address the frequency range of ~10 mHz – 10 Hz, as compared to the ~100 kHz for dynamic responses.[23] Here, the results reveal that the sample height periodically increases during BEPS cycling. Notably, the height peaks correspond to the maximum applied DC voltages (±40 V), with the height increase under -40 V being more pronounced than under +40 V.

We note that similar surface morphology changes have been observed in other ferroelectrics and electrochemical systems. Previously, this capability was used for the analysis of tip induced electrochemical transformation in oxides such as $LaAlO_3$-$SrTiO_3$ [24] and ceria[25] and Li-ion conductors.[26-30] Note that while in Li-ion conductors the nature of these tip induced electrochemical phenomena was reliably established as Li electroplating followed by oxidation (in air), in oxides it remained largely speculative, since obvious electrochemical mechanisms are absent. The concurrent Raman or time of flight secondary ion mass spectrometry (ToF-SIMS) studies were not able to identify the nature of the formed products due to extremely small volumes of generated products. For the nitrides explored here, qualitatively similar surface deformations are observed; the unusual aspect that has not been previously observed in oxide systems is the behavior of the remanent curve that still shows higher activity and polarization switching and the presence of fine structure on the hysteresis loop.

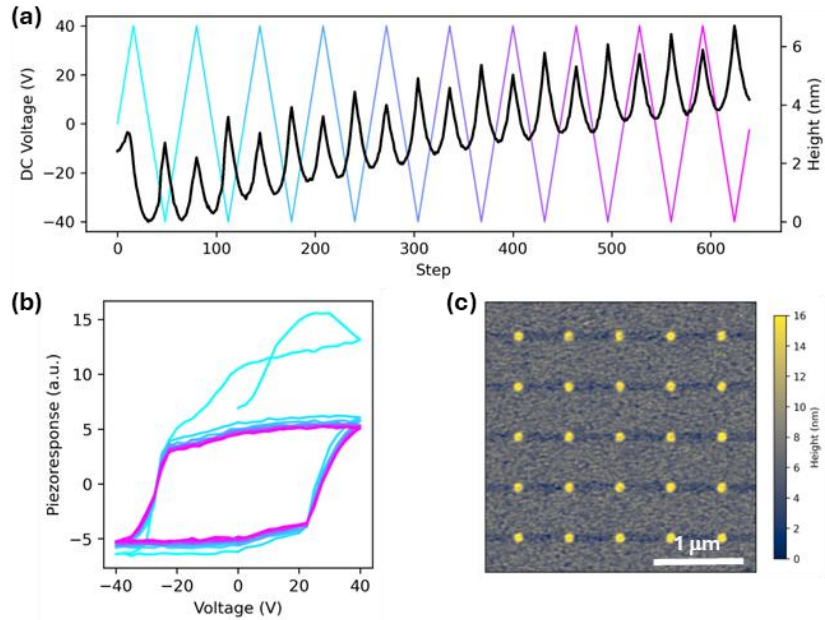

**Figure 1.** BEPS result of an $Al_{0.8}Sc_{0.2}N$ film in ambient. (a) The colorful curve is the DC waveform for BEPS measurement, where the color of loops represents measurement step; the black curve is the height evolution during the application of DC waveform. (b) BEPS hysteresis loop. (c) topography after BEPS measurement.

To get insight into the mechanisms of the observed phenomena and separate the anodic and cathodic processes, we have performed the BEPS measurements in the bipolar and unipolar first order reversal curve (FORC) mode.[31] The bipolar FORC mode consists of gradually increasing bipolar triangular waveforms (Figure 2a), while the unipolar FORC mode involves gradually increasing negative (Figure 2b) or positive (Figure 2c) waveforms. FORC measurements in ferroelectric materials allow for the assessment of switching process, similar approaches have also been utilized to investigate electrochemical activity on the surfaces of ionic conductors.

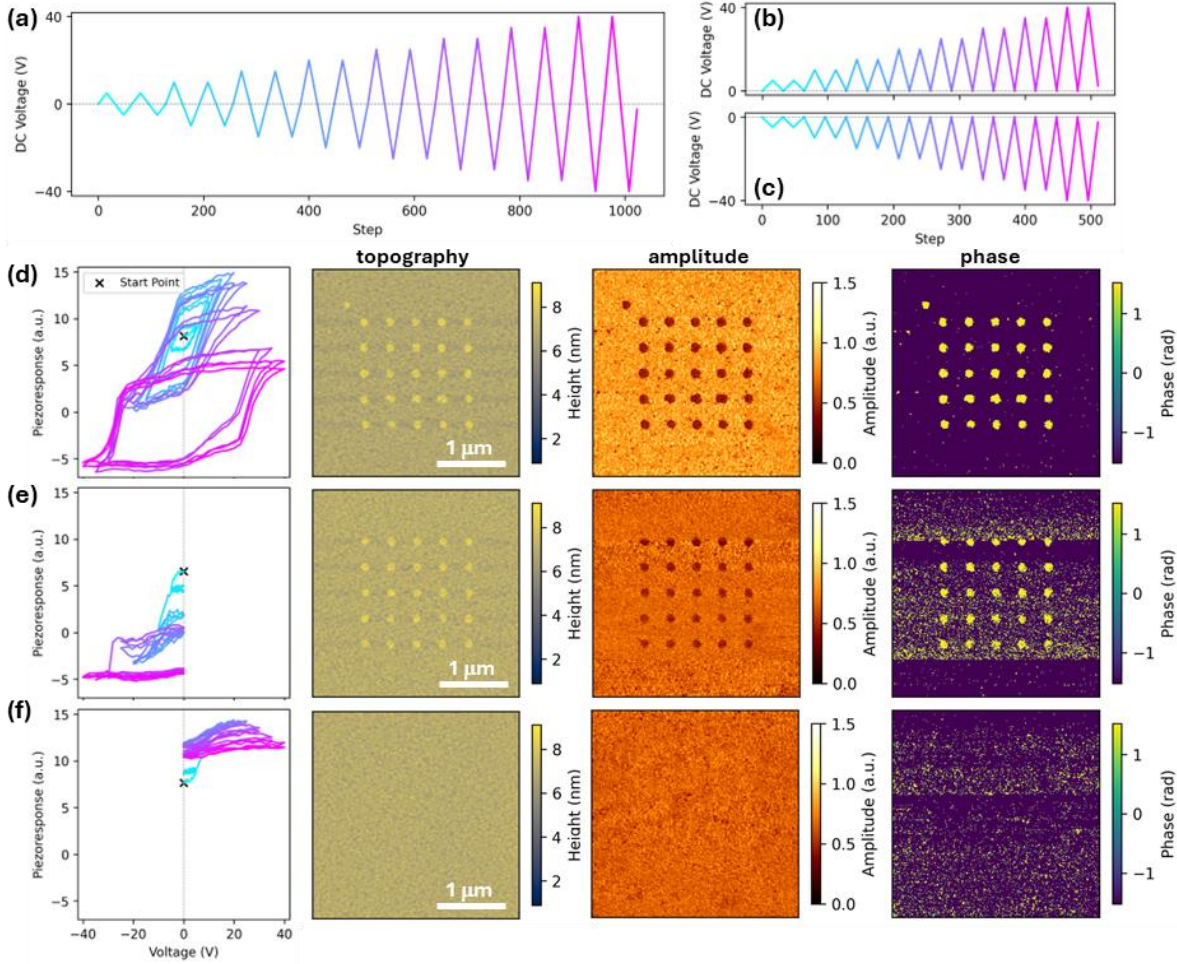

**Figure 2.** Polarity effect. (a) bi-polar, (b) negative unipolar, and (c) positive unipolar FORC waveform for BEPS measurement. d–f), BEPS loops, topography, BEPFM amplitude and phase after respective FORC-BEPS measurement. Here applications of unipolar FORC waveforms can reproduce the bi-polar FORC-BEPS loop at respective sides, as shown in Figure e and f.

Figures 2d-f show the BEPS loops measured under bipolar and unipolar FORC waveforms, along with the corresponding topography and BEPFM images following the application of the respective waveforms. The bipolar FORC BEPS results (Figure 2d) reveal a complex hysteresis loop structure, consisting of multiple loops that potentially represent multiple polarization switching stages. A key feature is the presence of small loops above larger loops. When the applied DC voltage exceeds ±20 V, the small loops rapidly open, and the center of the loops shifts downward. The topography measurements after bipolar FORC BEPS reveal the formation of a protruding structure on the sample surface. Figures 2e and 2f display the unipolar FORC BEPS results under negative and positive DC waveforms, respectively. Although repetitive polarization switching does not occur in these cases, the BEPS results for both negative and positive unipolar FORC waveforms partially replicate the transition process observed in bipolar FORC BEPS, specifically the progression from small loops to large loops. Notably, the negative unipolar FORC induces the formation of a protruding structure on the sample surface, as revealed by the topography measurements. In contrast, positive unipolar FORC does not result in significant morphological changes.

This behavior can be interpreted as the presence of the frozen polarization component on the free surface. We note that while PFM measurements can in principle be affected by multiple artefacts such as surface charging and electrostatic force effects,[32-38] contributions from the cantilever bending responses,[39, 40] and non-idealities in lock-in detection, these behaviors affect all voltage regimes equally and cannot lead to the behaviors observed in Figure 2. Hence, the data suggest the strong frozen polarization component for initial voltage range. On cycling and particularly voltage excursion towards negative polarities this frozen component disappears, and material regains (almost) classical switching behavior with the hysteresis loops spanning symmetric (+P, -P) domain. However, the anomaly on the loop remains. While we defer the discussion of possible mechanism to the later part of the text, it is worth mentioning that this behaviors cannot be observed in the macroscopic polarization–electric field measurements, since the latter do not provide information on frozen polarization components.

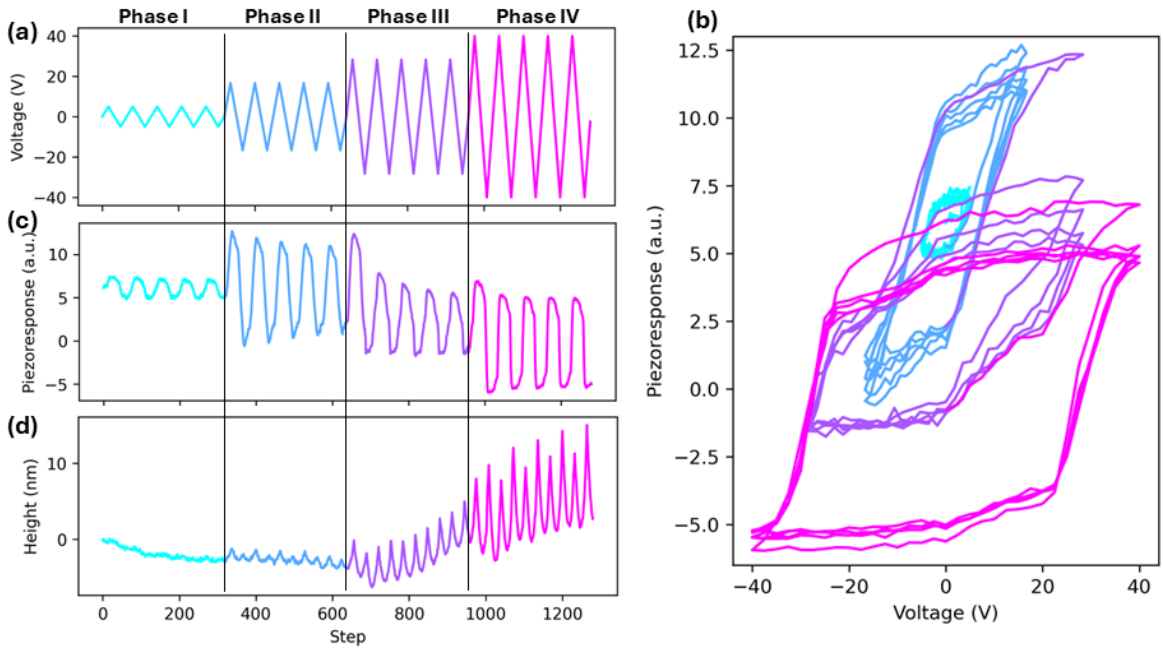

**Figure 3.** Multiple pseudo stable polarization states. (a) FORC waveform with multiple repetitions of intermediate states, the entire waveform is separated into four phases from I-IV. (b) FORC-BEPS hysteresis loops show multiple intermediate loops which are stable in respective phases of DC waveform. (c) Corresponding height evolution: Phase I→height gradually decreases; Phase II→height increases when the applied DC (either positive or negative) increases, height increase is larger under positive DC; Phase III→ height increases equally when the applied DC (either positive or negative) increases; Phase IV→height increase is larger under negative DC.

We have further explored the evolution of the surface topography and the coupling between the topographic changes and electromechanical response. Here, we applied a FORC waveform with multiple repetitions of each voltage cycles across four distinct voltage conditions (phase I—IV), as shown in Figure 3a, the corresponding BEPS hysteresis loops are shown in Figure 3b. The full loops indicate four distinct states. Phase I shows very small loops, the loops open at Phase II when the voltage increases. Subsequently, the loops shift in Phase III and further open in Phase IV. Note that loops are stable within the respective phases of voltage conditions. Correspondingly, the piezoresponse evolution in response to the FORC waveform is shown in Figure 3c: small amplitude changes in Phase I progressively increase in Phase II, amplitude in both Phase I and II are positive; subsequently, amplitude changes shift to the negative side in Phase III, ultimately amplitude is centered at 0 in Phase IV.

We also tracked the evolution of the sample height in response to the FORC waveform. In Phase I, the height exhibits a slight decrease as the applied voltage cycles, but no periodic modulations are visible. Phase II shows periodic height fluctuations that correspond directly to the voltage cycles; within each voltage cycle, two distinct height fluctuation peaks are observed, corresponding to the maximum DC voltages. In Phase III, the height fluctuation increases further due to the increasing DC voltage; during this phase, the fluctuations under maximum positive and negative DC voltages become similar in magnitude. Moving into Phase IV, the height fluctuations continue to increase. Additionally, Phases III and IV reveal a continuous overall increase in the height, indicating cumulative topographic changes in response to the applied voltage; in contrast, in Phase II there is no overall height changes, suggesting a reversible phenomenon.

These observations suggest a very complex picture of the bias-induced responses on the nitride surface. First, we note that these dynamics cannot be attributed to a single crosstalk mechanism, as was often the case for the PFM studies on the non-classical ferroelectrics. Secondly, the sensitivity for the topographic changes in the spectroscopic channel is of the order of 0.1 A. As such, we in principle expect to see topographic changes associated with the removal of a single N layer on the material surface and direct field strain coupling. As a relevant comparison, the lattice parameter of AlN is $c = 4.98$ Å, the average corrugation amplitude of the Al-N layers is 2.49 Å and piezoelectric constant is 5.5 pm/V, corresponding to displacements of the order of ~1 Å for applied 20 V bias. With these estimates being given, we also note that in the presence of multiple coupled mechanisms we expect PFM to be quantitative, but not selective. In other words, the potential applied to the probe partitions over multiple electromechanical subsystems in the tip-surface junction and measured response is the sum of response of each subsystem.[41] With this being said, the phase I and II of switching are clearly associated with the reversible dynamics that can be either intrinsic polarization switching or reversible surface electrochemistry. Transition to Phase III and IV is a clear onset to bulk switching and potentially concurrent electrochemical reaction.

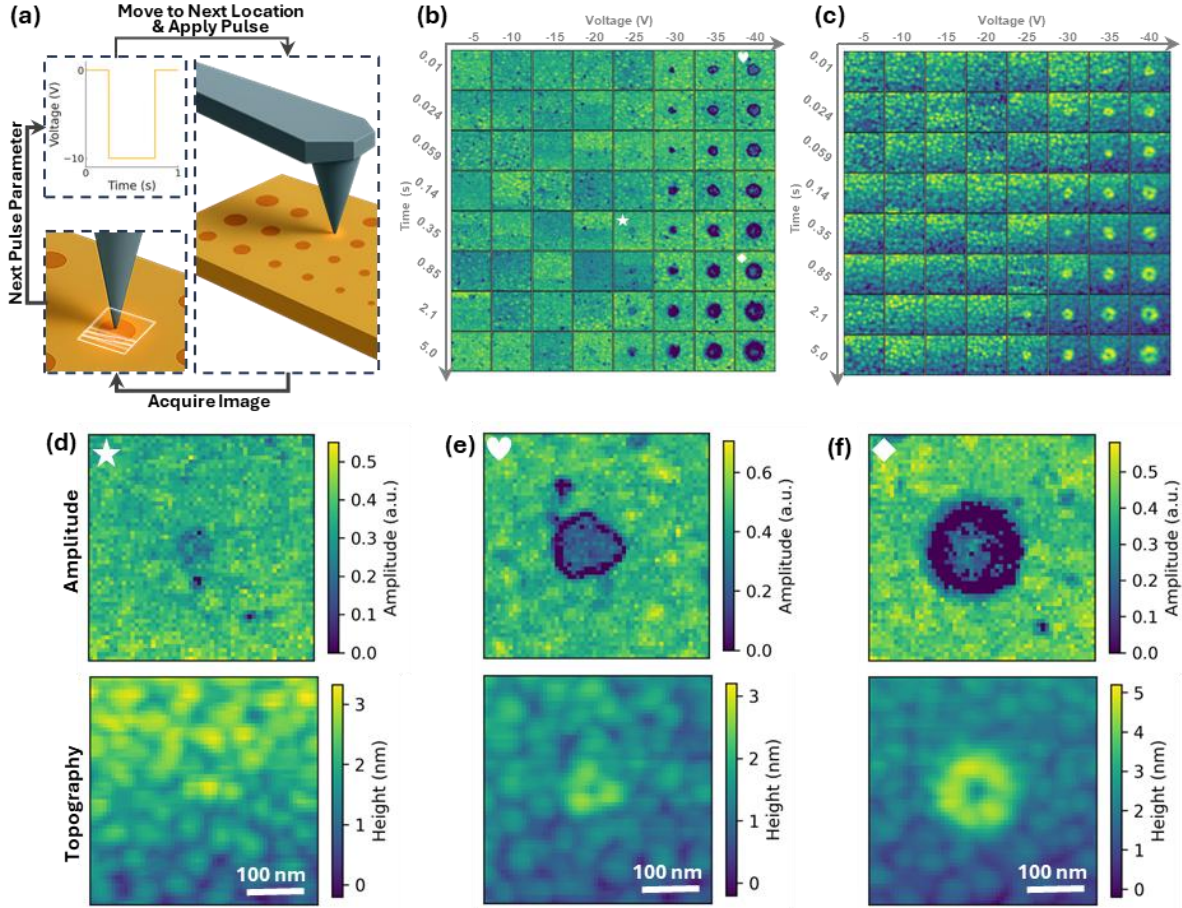

**Figure 4.** High-throughput domain writing experiments. (a) Process of domain writing experiment, where a DC pulse is applied to reverse the polarization and subsequently a PFM image measurement reveals the resultant domain created by the DC pulse; varying the DC pulse parameters allows to explore the domain growth procedure. (b) PFM amplitude results of high-throughput domain writing, where 8 pulse magnitude and 8 pulse duration conditions are explored (64 pulse conditions). (c) Topography results of high-throughput domain writing experiment. (d-f) highlight a few representative results, the corresponding results in (b-c) are marked as star or heart.

To get additional insight into the nature of these phenomena, we further explore the domain growth dynamics and correlative topography changes as a function of the bias pulse parameters. The as-grown $Al_{0.8}Sc_{0.2}N$ film is down-polarized, the polarization can be reversed by applying a negative direct current (DC) pulse via a scanning probe microscopy tip, and BEPFM mappings can reveal the ferroelectric domains induced by the DC pulse. We employed a high throughout domain writing and imaging approach based on AEcroscopy[42]—a platform for automated and autonomous scanning probe microscopy, enabling experiments that are difficult or unfeasible using traditional methods[43-46]—to explore the parameters spaces of pulse conditions (i.e., pulse magnitude and duration). As shown in Figure 4a, this experiment starts with applying a DC pulse at the center of the measurement area to switch the polarization (whether the polarization can be reversed based on the magnitude and duration of the applied DC pulse), followed by a high resolution BE PFM image measurement to map the polarization state of this area. Then, the same measurement is performed at the next area by applying a different DC pulse. Here, using AEcroscopy platform, we can change the pulse conditions and perform BEPFM mapping automatically with a workflow script, allowing us to systematically explore the DC pulse conditions to study the effect of DC pulse parameters on polarization switching and topography change.

Figure 4b shows the BE PFM amplitude maps, each square represents independent BEPFM maps after applying a spectrum of pulses ranging from – (5 – 40) V and 0.01 – 5 s. Figure 4c shows the corresponding topography maps. In the amplitude maps, we observed domain switching when the applied DC pulse is large and long. The first domain appears at Vdc = -25 V and t = 0.35 s, which is marked by a star; meanwhile, topography also shows a slight morphology change. The domain progressively becomes larger when DC pulse increases, along with more pronounced topography changes. Several amplitude and topography maps showing representative domain structures and morphology changes are enlarged in Figure 4d-f.

Upon careful examination of the data in Figure 4, several trends are becoming obvious. The low voltage regimes are associated with no domain formation (corresponding to Phase I on Figure 3) and formation of the shadow domains, meaning the PFM domains with weakly changing contrast and incomplete polarization switching. In these regimes, there is no topographic changes within the image. At the same time, for higher biases both domain switching and formation of strong ferroelectric domains and the topographic deformations are observed. The careful exploration of the boundary between the two suggests that polarization switching precedes the surface deformation by ~5 V, as evidenced by several instances where strong domain has formed and topography did not change, formation of strong domains away from the contact area, and generally smaller size of region with deformation compared to the domain. It is also remarkable that the surface deformation patterns are clearly aligned with the grains on the sample surface, with the integer multiple of grains undergoing expansion process. We further note that deformed regions have a characteristic "doughnut" shape, suggesting that the tip force is significantly large to make forming material flow around the tip. The corollary to this observation is that the reaction product is relatively soft compared to AlN, and that the surface deformations measured by the height spectroscopy will be well below the actual surface deformation.

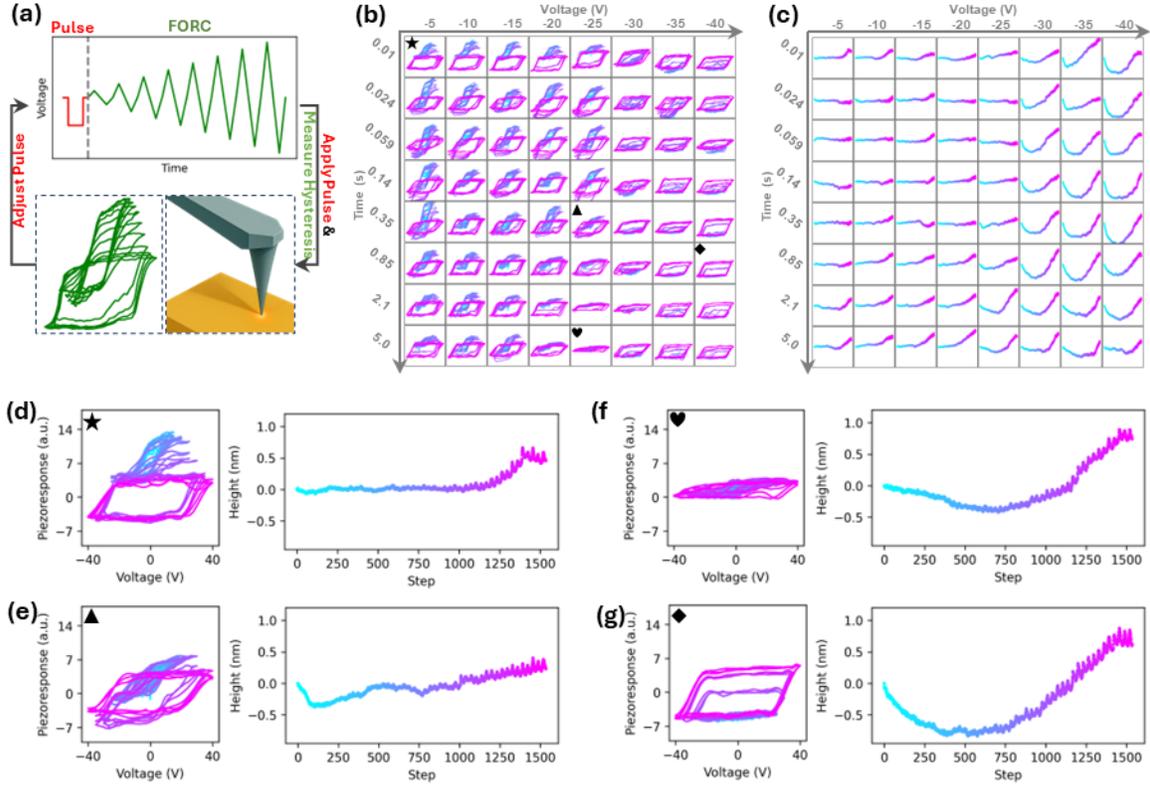

**Figure 5.** Set pulse (SP)-FORC-BEPS hysteresis measurement. (a) Process of SP-FORC-BEPS measurement, where a set pulse is applied before the FORC-BEPS measurement, the set pulse parameters are consistent with the parameters for domain writing experiment. (b) high-throughput exploration of set pulse effects on hysteresis loops, (c) the profile showing height evolution during the FORC-BEPS measurement, which indicate the effect of set-pulse on the height evolution under FORC waveform. (d-g) a few representative loops are highlighted and amplified, the loops and profiles are marked using star, triangle, heart, and diamond in (b). The small loops are almost closed if the pre set-pulses reverse the polarization, correspondingly height change is more significant.

We further designed an automated experiment incorporating a DC set-pulse prior to the FORC waveform for BEPS measurements, as shown in Figure 5a. This experiment was aimed at exploring how pre-set conditions influence polarization dynamics. In other words, while FORC measurements slowly change the surface conditions via outer envelope voltage and provide the differential detection of induced changes via internal sweeps, the preset pulse measurements explore the effect of a single conditioning pulse parameters on subsequent dynamics. Here, the set-pulse was applied and modified automatically using our AEcroscopy platform. The set-pulse array used for BEPS measurement was identical to the DC pulse array in the previous domain writing experiments. The results of the set-pulse FORC-BEPS experiment are shown in Figure 5b, where the hysteresis loops exhibit multiple stages in each measurement. It is observed that as the set-pulse magnitude and duration increase, the small hysteresis loops progressively shift from the top side to the bottom side within the larger hysteresis loops. The evolution of sample height during the application of the FORC waveform was also monitored, with the results displayed in Figure 5c. These data show that topographic changes become increasingly pronounced when the pre-set pulses reverse the polarization state.

Figures 5d-g show some representative hysteresis loops and corresponding height evolution profiles. In Figure 5d, when the set-pulse is small, the hysteresis loop is similar to the loop measured from the pristine film (Figure 2d). The height evolution profile indicates only a height increase at the end. In Figure 5e, when the set-pulse begins to alter the polarization condition (see corresponding domain structure in Figure 4), the small loop starts to shift toward downside. The corresponding height profile shows an initial rapid decrease followed by a continuous increase. In Figure 5f, when the pre-set pulse is slightly longer than the duration required to induce a polarization state change, a collapse of the hysteresis loops is observed. The height profile in this case reveals a slow decrease in the first half of the measurement, followed by a more rapid increase in the second half. In Figure 5g, when the set-pulse fully reverses the polarization before the FORC-BEPS measurement (see corresponding domain structure in Figure 4), the small loop shifts entirely to the bottom side and tends to collapse. The height profile here shows a more pronounced decrease during the first half and an increase during the second half. These results indicate that the polarization dynamics and surface morphology changes are closely related to the pre-set.

The common behavior we observe is that formation of smaller loops is associated with the small topographic changes at initial voltages and subsequent surface expansion. Sufficiently strong preset pulses remove the small loop effects, and the topographic evolution during the loop opening follows compression and subsequent expansion pattern. In fact, under certain combinations of parameters (Figure 5f) the responses from two ferroelectric sub-systems can compensate each other and the resulting hysteresis loops are almost closed.

To establish the nature of the surface reaction products, we have attempted ToF-SIMS. However, the results were below detection limit of the technique, as can be expected given the very small volumes of the modified regions on the sample surface. We note that while large-scale electrochemical reactions can provide information on the product, we expect it to be simple field-induced hydrolysis of AlN with the formation of the volatile N-containing species and non-volatile $Al(OH)_3$. Of interest toward the understanding the reaction mechanisms are the early stage of this process associate with effects localized both laterally to several tens of nanometers and vertically to the surface layer, and for these the sensitivity on detection limits of PFM imaging and spectroscopies are many orders of magnitude below that of chemical imaging methods.

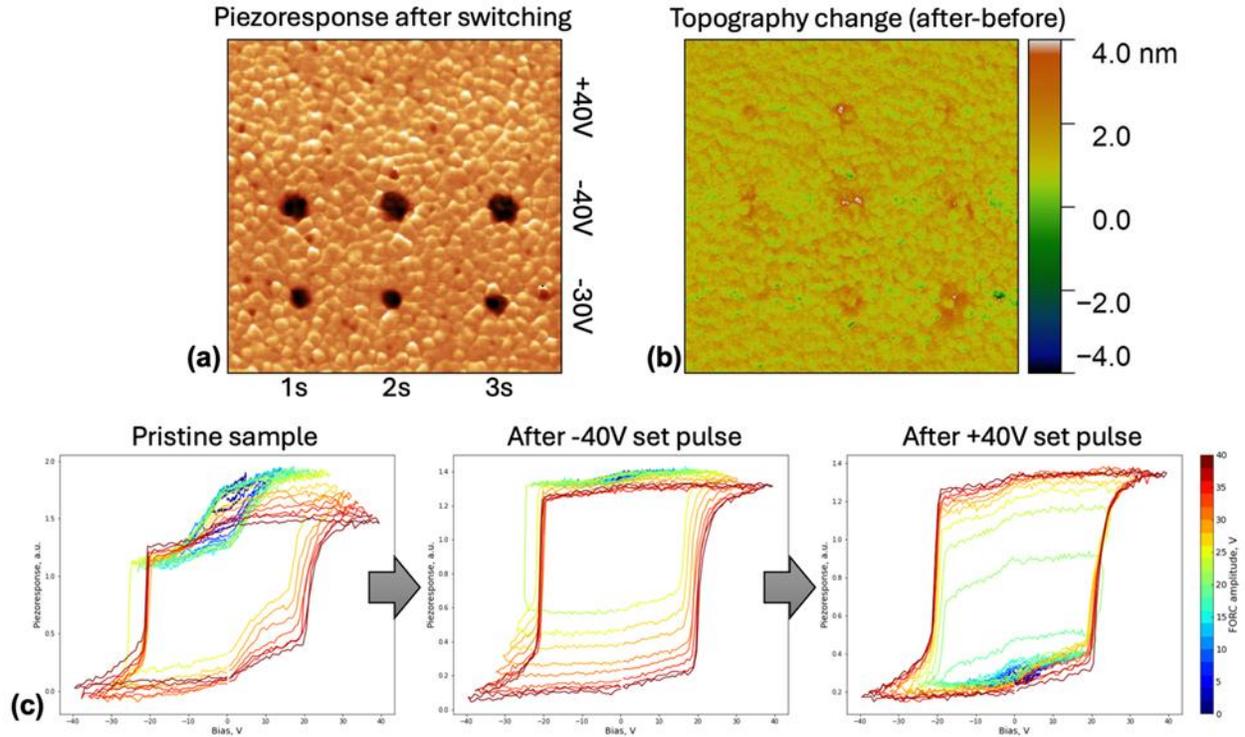

**Figure 6.** Reproducing the domain writing and the set-pulse hysteresis loops experiment in vacuum. (a) domain writing experiments. (b) A very small topography change is seen in vacuum, under both positive and negative pulses. (c) the set-pulse also alters the location of small loop; in addition, if the set pulse reverses the polarization, the small loop tends to collapse.

Hence to investigate the potential influence of surface electrochemistry due to the ambient environment, we conducted domain writing and set-pulse BEPS experiments in a vacuum environment. Since these vacuum experiments require manual operation by a human operator, we only tested a few sets of parameters. Figures 6a and 6b show the piezoresponse and height change maps after the application of several representative DC pulses, the magnitude and duration of the DC pulses are indicated in Figure 6a. The piezoresponse maps in Figure 6a show domains indicating polarization reversal under DC pulses of -30 V and -40 V, with the domain size increasing as the pulse duration lengthens. In contrast, applying a +40 V DC pulse does not result in polarization reversal. The corresponding morphological changes induced by these DC pulses are shown in Figure 6b. While +40 V does not reverse the polarization, it also induces small morphology changes, this is consistent with the observations in the height tracking profile shown in Figure 3, where both positive and negative voltages caused height changes.

In the set-pulse BEPS measurement shown in Figure 6c, we observed double hysteresis loops in the pristine film. When a set-pulse of -40 V was applied to reverse the polarization before the FORC BEPS measurement, the small loops shifted to the downside of the hysteresis loop. Subsequently, the small loops shifted back to the topside when the polarization was reversed to the same state as the pristine film by applying a +40 V set-pulse. However, after repeated cycling of polarization, the small loops tended to collapse. These observations in vacuum align with previous results obtained under ambient conditions, suggesting that the observed phenomena are primarily due to the intrinsic properties of $Al_{0.8}Sc_{0.2}N$ rather than being influenced by the environmental factors.

Based on the PFM observations, we note that some aspects of the PFM polarization switching in $Al_{0.8}Sc_{0.2}N$ studied here are very similar to the previously reported results on the $Zn_{1-x}Mg_xO$ system,[8] including the formation of the minor and major loops and the shadow domains. However, the element that it absents in the $Zn_{1-x}Mg_xO$ and oxide fluorites is the persistent surface deformation for high biases. For $Zn_{1-x}Mg_xO$, we postulated the presence of the shark teeth domains at the first stage of switching, corresponding to partial switching that will not be readily detectable in P-E measurements but will strongly affect the electromechanical responses of the surface. While such domains in our knowledge have not been observed for $Zn_{1-x}Mg_xO$, they have been visualized via STEM measurements for nitrides similar to the one studied here.[16, 18] Based on the commonalities in PFM studies for oxide and nitride wurtzite and observation of these partial switching domains for the nitrides, we conclude that this mechanism is likely common for the wurtzite ferroelectrics.

We further discuss the possible mechanisms behind the low voltage dynamics in PFM observed here. Based on PFM studies, we conclude that it corresponds to the stages before the bulk switching and can be associated with the formation of single shark tooth domain and surface electrochemical changes. We further note that polarization switching in nitrides is associated with the changes of surface termination, creating opportune conditions for electrochemical reactivity, especially at the moving domain wall.

We performed the phase field modelling of the polarization switching to explore the mechanisms. A possible theoretical explanation uses the assumption that polar and dielectric properties of the electrically open surface of wurtzite ferroelectric $Al_{0.8}Sc_{0.2}N$ undergoes significant changes due to the adsorption of ions and/or vacancies during electrochemical reactions, which can take place under the biased tip of the PFM probe. In result of the changes an ultrathin gradient-type layer appears at the surface and acts as the source of screening charges, being the source of domain formation due to incomplete screening. Figure 7 is calculated by the finite element modelling (FEM) in the framework of Landau-Ginzburg-Devonshire (LGD) approach (see details in Supplementary Information). The bias applied between the probe tip apex and bottom electrode changes in time periodically, its amplitude increases linearly in time as shown in Figure 7a. Figures 7b-7c show the time dependences of the average polarization and surface displacement in response to the bias changes. A hysteresis-type dependence of the average polarization versus applied bias is shown in Figure 7d (see also Figure S1 for details). Snapshots of the domain pattern under the probe, which correspond to the moments of time 1 – 15, are shown in Figure 7e.

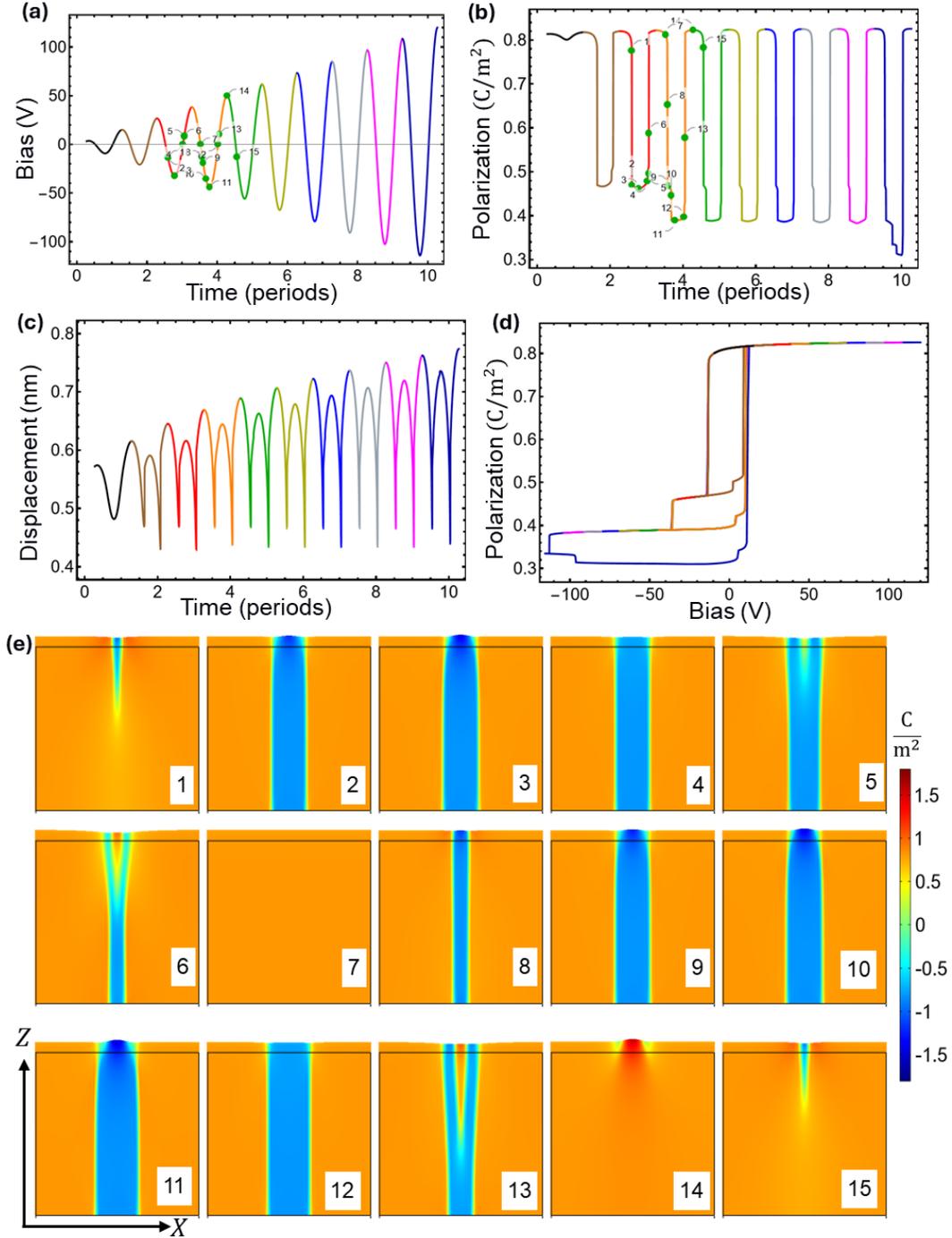

**Figure 7.** The time dependence of the bias applied between the probe tip apex and bottom electrode **(a)**, the average polarization **(b)**, the vertical displacement of the surface below the tip (the local piezoelectric response) **(c)**, and the bias dependence of the average polarization **(d)** calculated by FEM in the heterostructure "probe – $Al_{0.8}Sc_{0.2}N$ layer - electrode". The distribution of polarization **(e)** in the cross-section of the bilayer at the moments of time numbered from "1" to "15" shown by the pointers in (a) and (b). The total thickness of the $Al_{0.8}Sc_{0.2}N$ layer is 40 nm, the diffusion length $h_d$ of the gradient-type PE layer is equal to 1 nm; the tip-surface contact radius $R$ is 2 nm. LGD parameters and elastic constants are listed in **Tables SI-SII**.

As one can see from Figure 7b-7d, the small bias amplitude is insufficient to induce the local reversal of polarization under the probe. The strongly up-shifted polarization loop opens with an increase in the bias amplitude, which corresponds to the partial polarization switching. The domain under the probe reaches the bottom electrode and grows in the lateral directions for higher amplitudes of applied bias. However, the lateral growth of the domain does not lead to the polarization reversal of the entire computation cell; and this is the reason for the hysteresis loop vertical asymmetry. Every new cycle of applied bias (with higher amplitude) shifts the loop down at nearly equidistant value, but the loop remains strongly asymmetrical. One can see three large vertical steps in the figure, showing the process of the loop moving down is step-like. The small step-like features at the hysteresis loops are associated with the collapse of domain walls being similar to Barkhausen jumps. These jumps appear when the small domain, which is growing inside the larger domain, reaches the boundaries of the larger domain, as illustrated in Figure 7e.

However, results shown in Figure 7d do not describe all experimentally observed piezo-response loops, shown in Figure 2-3, and generally describes only later stages of switching. Based on this comparison, we attribute early polarization signatures to the factors that are not a part of phase field mode, namely surface chemistry. Our preliminary modelling, which assumes that the tip-surface contact radius $R$ increases with an increase in voltage amplitude due to the electro-chemically induced formation of conductive meniscus under the probe, shows that piezoresponse loop becomes almost vertically symmetrical with an increase in the bias amplitude; at that the process of loop symmetrization is step-like. This happens because the polarization switching of a ferroelectric layer placed in the flat capacitor corresponds to symmetrical loops. However, the loop does not "falling-down" as a whole with increase in $R$ (as shown in Figure 2-3). The falling-down effect may be related with the in-depth electrochemical changes of the $Al_{0.8}Sc_{0.2}N$ layer, when the thickness of the gradient-type PE layer increases strongly with the bias increase. In this case the total thickness of the dead layer increases leading to the decrease in the remanent polarization of the $Al_{0.8}Sc_{0.2}N$ layer due to the proximity effect.[47, 48]

In summary, our study investigated the polarization switching behavior in wurtzite-structure ferroelectric nitrides using piezoresponse force microscopy (PFM) under both ambient and vacuum conditions. We observed a multi-stage switching mechanism, with distinct contributions from changes in surface termination, electrochemical reactivity, and bulk ferroelectric switching. Remarkably, bulk switching in these materials initiates only after a surface transformation—presumably involving nitrogen loss and the formation of a reactive electrochemical layer—analogous in some respects to solid-electrolyte interphase (SEI) formation in batteries. While such surface-controlled switching pathways are unconventional in classical ferroelectric frameworks, they find parallels in materials such as ceria, lanthanum aluminates, and strontium titanate, where water-mediated redox cycles[49] play a critical role in conductivity and field-driven processes.

Our results further demonstrate the power of advanced PFM techniques in probing these phenomena. In particular, the combination of electromechanical response mapping with high-resolution topographic imaging enables a unique view into the dynamics of individual screening charges, the onset of local electrochemical reactions, and strain-coupled domain evolution at the single-unit-cell level. Although lacking direct chemical specificity, these measurements access spatial and temporal regimes beyond the reach of conventional spectroscopic tools and spanning the diffusion times of multiple experimentally relevant ions, providing essential insights into the complex, bias-induced processes that govern polarization dynamics in these systems.

Finally, while our experiments focus on free surfaces, the physics we reveal likely extend to buried interfaces as well. In both cases, the requirement for surface or interface termination change appears to be a prerequisite for switching. On free surfaces, adsorbed water layers and ambient chemical species play an active role in this transformation, highlighting the critical importance of surface conditioning and environmental control. We hypothesize that similar electrochemically mediated switching mechanisms will operate at buried interfaces in device environments. We therefore anticipate that future studies combining PFM with chemical mapping techniques and in situ STEM will be key to uncovering and ultimately engineering interface-driven switching in nitride ferroelectrics.

**Acknowledgements**
This effort (PFM and ToF-SIMS experiments) was supported by the Center for Nanophase Materials Sciences (CNMS), which is a US Department of Energy, Office of Science User Facility at Oak Ridge National Laboratory, and using instrumentation within ORNL's Materials Characterization Core provided by UT-Battelle, LLC under Contract No. DE-AC05-00OR22725 with the U.S. Department of Energy. This research was sponsored by the INTERSECT Initiative as part of the Laboratory Directed Research and Development Program of Oak Ridge National Laboratory, managed by UT-Battelle, LLC for the US Department of Energy under contract DE-AC05-00OR22725. The work of A.N.M. is funded by the National Research Foundation of Ukraine (project "Silicon-compatible ferroelectric nanocomposites for electronics and sensors", grant N 2023.03/0127). The work of E.A.E. is funded by the National Research Foundation of Ukraine (project "Manyfold-degenerated metastable states of spontaneous polarization in nanoferroics: theory, experiment and perspectives for digital nanoelectronics", grant N 2023.03/0132). Obtained results were visualized in Mathematica 14.0[50]. SVK effort was supported by the UT Knoxville start-up funding. Film growth (H.F.) was supported by MEXT Program: Data Creation and Utilization Type Material Research and Development Project (No. JPMXP1122683430) and MEXT Initiative to Establish Next-generation Novel Integrated Circuits Centers (X-NICS) (JPJ011438), and the Japan Science and Technology Agency (JST) as part of Adopting Sustainable Partnerships for Innovative Research Ecosystem (ASPIRE), Grant Number JPMJAP2312.

**Author Contributions**
Y. L. conceived the project, designed and performed PFM experiments. S.V.K. proposed interpretation. E.A.E. and A.N.M performed simulation. A.V.I. performed vacuum PFM. H.F., N.S., and K.O. made the $Al_{0.8}Sc_{0.2}N$ sample and performed fundamental characterization of crystal structure and ferroelectricity. Y. L. and S.V.K. wrote the manuscript. All authors edited the manuscript.

**Conflict of Interest**
The authors confirm no conflicts of interest.

**References**
(1) Böscke, T.; Müller, J.; Bräuhaus, D.; Schröder, U.; Böttger, U. Ferroelectricity in hafnium oxide thin films. *Applied Physics Letters* **2011**, *99* (10).
(2) Fichtner, S.; Wolff, N.; Lofink, F.; Kienle, L.; Wagner, B. AlScN: A III-V semiconductor based ferroelectric. *Journal of Applied Physics* **2019**, *125* (11).
(3) Hayden, J.; Hossain, M. D.; Xiong, Y.; Ferri, K.; Zhu, W.; Imperatore, M. V.; Giebink, N.; Trolier-McKinstry, S.; Dabo, I.; Maria, J.-P. Ferroelectricity in boron-substituted aluminum nitride thin films. *Physical Review Materials* **2021**, *5* (4), 044412.
(4) Ferri, K.; Bachu, S.; Zhu, W.; Imperatore, M.; Hayden, J.; Alem, N.; Giebink, N.; Trolier-McKinstry, S.; Maria, J.-P. Ferroelectrics everywhere: Ferroelectricity in magnesium substituted zinc oxide thin films. *Journal of Applied Physics* **2021**, *130* (4).
(5) Zhou, D.; Xu, J.; Li, Q.; Guan, Y.; Cao, F.; Dong, X.; Müller, J.; Schenk, T.; Schröder, U. Wake-up effects in Si-doped hafnium oxide ferroelectric thin films. *Applied Physics Letters* **2013**, *103* (19).
(6) Casamento, J.; Baksa, S. M.; Behrendt, D.; Calderon, S.; Goodling, D.; Hayden, J.; He, F.; Jacques, L.; Lee, S. H.; Smith, W. Perspectives and progress on wurtzite ferroelectrics: Synthesis, characterization, theory, and device applications. *Applied Physics Letters* **2024**, *124* (8).
(7) Yazawa, K.; Hayden, J.; Maria, J.-P.; Zhu, W.; Trolier-McKinstry, S.; Zakutayev, A.; Brennecka, G. L. Anomalously abrupt switching of wurtzite-structured ferroelectrics: simultaneous non-linear nucleation and growth model. *Materials Horizons* **2023**, *10* (8), 2936-2944.
(8) Yang, J.; Ievlev, A. V.; Morozovska, A. N.; Eliseev, E. A.; Poplawsky, J. D.; Goodling, D.; Spurling, R. J.; Maria, J. P.; Kalinin, S. V.; Liu, Y. Coexistence and Interplay of Two Ferroelectric Mechanisms in $Zn_{1-x}Mg_xO$. *Advanced Materials* **2024**, *36* (39), 2404925.
(9) Behrendt, D.; Samanta, A.; Rappe, A. M. Ferroelectric Fractals: Switching Mechanism of Wurtzite AlN. *arXiv preprint arXiv:2410.18816* **2024**.
(10) Lee, C.-W.; Yazawa, K.; Zakutayev, A.; Brennecka, G. L.; Gorai, P. Switching it up: New mechanisms revealed in wurtzite-type ferroelectrics. *Science Advances* **2024**, *10* (20), eadl0848.
(11) Fichtner, S.; Schönweger, G.; Lee, C.-W.; Yazawa, K.; Gorai, P.; Brennecka, G. L. Polarization and domains in wurtzite ferroelectrics: Fundamentals and applications. *Applied Physics Reviews* **2025**, *12* (2).
(12) Yang, F. Physics of Ferroelectric Wurtzite $Al_{1-x}Sc_xN$ Thin Films. *Advanced Electronic Materials* **2025**, *11* (2), 2400279.
(13) Kalinin, S. V.; Eliseev, E. A.; Morozovska, A. N. Adsorption of ions from aqueous solutions by ferroelectric nanoparticles. *Physical Review Applied* **2025**, *23* (1), 014081.
(14) Kalinin, S. V.; Bonnell, D. A.; Alvarez, T.; Lei, X.; Hu, Z.; Ferris, J.; Zhang, Q.; Dunn, S. Atomic polarization and local reactivity on ferroelectric surfaces: a new route toward complex nanostructures. *Nano Letters* **2002**, *2* (6), 589-593.
(15) Kakekhani, A.; Ismail-Beigi, S. Polarization-driven catalysis via ferroelectric oxide surfaces. *Physical Chemistry Chemical Physics* **2016**, *18* (29), 19676-19695.
(16) Wolff, N.; Grieb, T.; Schönweger, G.; Krause, F. F.; Streicher, I.; Leone, S.; Rosenauer, A.; Fichtner, S.; Kienle, L. Electric field-induced domain structures in ferroelectric AlScN thin films. *Journal of Applied Physics* **2025**, *137* (8).
(17) Calderon, S.; Hayden, J.; Baksa, S. M.; Tzou, W.; Trolier-McKinstry, S.; Dabo, I.; Maria, J.-P.; Dickey, E. C. Atomic-scale polarization switching in wurtzite ferroelectrics. *Science* **2023**, *380* (6649), 1034-1038.
(18) Wolff, N.; Schönweger, G.; Streicher, I.; Islam, M. R.; Braun, N.; Straňák, P.; Kirste, L.; Prescher, M.; Lotnyk, A.; Kohlstedt, H. Demonstration and STEM analysis of ferroelectric switching in MOCVD-grown single crystalline $Al_{0.85}Sc_{0.15}N$. *Advanced Physics Research* **2024**, *3* (5), 2300113.

(19) Liu, Y.; Ievlev, A.; Casamento, J.; Hayden, J.; Trolier-McKinstry, S.; Maria, J. P.; Kalinin, S. V.; Kelley, K. P. The Interplay Between Ferroelectricity and Electrochemical Reactivity on the Surface of Binary Ferroelectric $Al_xB_{1-x}N$. *Advanced Electronic Materials* **2024**, *10* (2), 2300489.

(20) Sun, N.; Okamoto, K.; Yasuoka, S.; Doko, S.; Matsui, N.; Irisawa, T.; Tsunekawa, K.; Katase, T.; Koganezawa, T.; Nakatani, T. High stability of the ferroelectricity against hydrogen gas in (Al, Sc) N thin films. *Applied Physics Letters* **2024**, *125* (3).

(21) Yasuoka, S.; Shimizu, T.; Tateyama, A.; Uehara, M.; Yamada, H.; Akiyama, M.; Hiranaga, Y.; Cho, Y.; Funakubo, H. Effects of deposition conditions on the ferroelectric properties of $(Al_{1-x}Sc_x)$ N thin films. *Journal of Applied Physics* **2020**, *128* (11).

(22) Jesse, S.; Maksymovych, P.; Kalinin, S. V. Rapid multidimensional data acquisition in scanning probe microscopy applied to local polarization dynamics and voltage dependent contact mechanics. *Applied Physics Letters* **2008**, *93* (11), 112903, Article. DOI: Artn 112903

10.1063/1.2980031.

(23) Morozovska, A. N.; Eliseev, E. A.; Kalinin, S. V. Electrochemical strain microscopy with blocking electrodes: The role of electromigration and diffusion. *Journal of Applied Physics* **2012**, *111* (1), Article. DOI: Artn 014114

10.1063/1.3675508.

(24) Kumar, A.; Arruda, T. M.; Kim, Y.; Ivanov, I. N.; Jesse, S.; Bark, C. W.; Bristowe, N. C.; Artacho, E.; Littlewood, P. B.; Eom, C. B.; et al. Probing surface and bulk electrochemical processes on the $LaAlO_3$-$SrTiO_3$ interface. *ACS Nano* **2012**, *6* (5), 3841-3852, Article. DOI: 10.1021/nn204960c From NLM PubMed-not-MEDLINE.

(25) Yang, N.; Doria, S.; Kumar, A.; Jang, J. H.; Arruda, T. M.; Tebano, A.; Jesse, S.; Ivanov, I. N.; Baddorf, A. P.; Strelcov, E. Water-mediated electrochemical nano-writing on thin ceria films. *Nanotechnology* **2014**, *25* (7), 075701.

(26) Kalinin, S.; Balke, N.; Jesse, S.; Tselev, A.; Kumar, A.; Arruda, T. M.; Guo, S. L.; Proksch, R. Li-ion dynamics and reactivity on the nanoscale. *Materials Today* **2011**, *14* (11), 548-558, Review. DOI: 10.1016/S1369-7021(11)70280-2.

(27) Arruda, T. M.; Kumar, A.; Jesse, S.; Veith, G. M.; Tselev, A.; Baddorf, A. P.; Balke, N.; Kalinin, S. V. Toward quantitative electrochemical measurements on the nanoscale by scanning probe microscopy: environmental and current spreading effects. *ACS Nano* **2013**, *7* (9), 8175-8182, Article. DOI: 10.1021/nn4034772 From NLM Medline.

(28) Kumar, A.; Chen, C.; Arruda, T. M.; Jesse, S.; Ciucci, F.; Kalinin, S. V. Frequency spectroscopy of irreversible electrochemical nucleation kinetics on the nanoscale. *Nanoscale* **2013**, *5* (23), 11964-11970. DOI: 10.1039/c3nr03953f From NLM PubMed-not-MEDLINE.

(29) Kumar, A.; Arruda, T. M.; Tselev, A.; Ivanov, I. N.; Lawton, J. S.; Zawodzinski, T. A.; Butyaev, O.; Zayats, S.; Jesse, S.; Kalinin, S. V. Nanometer-scale mapping of irreversible electrochemical nucleation processes on solid Li-ion electrolytes. *Sci Rep* **2013**, *3* (1), 1621. DOI: 10.1038/srep01621 From NLM Medline.

(30) Jesse, S.; Kumar, A.; Arruda, T. M.; Kim, Y.; Kalinin, S. V.; Ciucci, F. Electrochemical strain microscopy: Probing ionic and electrochemical phenomena in solids at the nanometer level. *Mrs Bulletin* **2012**, *37* (7), 651-658, Article. DOI: 10.1557/mrs.2012.144.

(31) Strelcov, E.; Belianinov, A.; Hsieh, Y. H.; Jesse, S.; Baddorf, A. P.; Chu, Y. H.; Kalinin, S. V. Deep data analysis of conductive phenomena on complex oxide interfaces: physics from data mining. *ACS Nano* **2014**, *8* (6), 6449-6457, Article. DOI: 10.1021/nn502029b From NLM PubMed-not-MEDLINE.

(32) Qiao, H.; Seol, D.; Sun, C.; Kim, Y. Electrostatic contribution to hysteresis loop in piezoresponse force microscopy. *Applied Physics Letters* **2019**, *114* (15). DOI: Artn 152901

10.1063/1.5090591.
(33) Gomez, A.; Puig, T.; Obradors, X. Diminish electrostatic in piezoresponse force microscopy through longer or ultra-stiff tips. *Applied Surface Science* **2018**, *439*, 577-582.
(34) Kim, S.; Seol, D.; Lu, X.; Alexe, M.; Kim, Y. Electrostatic-free piezoresponse force microscopy. *Sci Rep* **2017**, *7* (1), 41657. DOI: 10.1038/srep41657 From NLM PubMed-not-MEDLINE.
(35) Seol, D.; Kim, B.; Kim, Y. Non-piezoelectric effects in piezoresponse force microscopy. *Current Applied Physics* **2017**, *17* (5), 661-674. DOI: 10.1016/j.cap.2016.12.012.
(36) Balke, N.; Maksymovych, P.; Jesse, S.; Herklotz, A.; Tselev, A.; Eom, C. B.; Kravchenko, II; Yu, P.; Kalinin, S. V. Differentiating Ferroelectric and Nonferroelectric Electromechanical Effects with Scanning Probe Microscopy. *ACS Nano* **2015**, *9* (6), 6484-6492. DOI: 10.1021/acsnano.5b02227 From NLM PubMed-not-MEDLINE.
(37) Eliseev, E. A.; Morozovska, A. N.; Ievlev, A. V.; Balke, N.; Maksymovych, P.; Tselev, A.; Kalinin, S. V. Electrostrictive and electrostatic responses in contact mode voltage modulated scanning probe microscopies. *Applied Physics Letters* **2014**, *104* (23), Article. DOI: 10.1063/1.4882861.
(38) Balke, N.; Maksymovych, P.; Jesse, S.; Kravchenko, II; Li, Q.; Kalinin, S. V. Exploring local electrostatic effects with scanning probe microscopy: implications for piezoresponse force microscopy and triboelectricity. *ACS Nano* **2014**, *8* (10), 10229-10236. DOI: 10.1021/nn505176a From NLM PubMed-not-MEDLINE.
(39) Huey, B. D.; Ramanujan, C.; Bobji, M.; Blendell, J.; White, G.; Szoszkiewicz, R.; Kulik, A. The importance of distributed loading and cantilever angle in piezo-force microscopy. *Journal of Electroceramics* **2004**, *13* (1-3), 287-291, Article; Proceedings Paper. DOI: DOI 10.1007/s10832-004-5114-y.
(40) Jesse, S.; Baddorf, A. P.; Kalinin, S. V. Dynamic behaviour in piezoresponse force microscopy. *Nanotechnology* **2006**, *17* (6), 1615-1628. DOI: 10.1088/0957-4484/17/6/014 From NLM PubMed-not-MEDLINE.
(41) Kalinin, S. V.; Bonnell, D. A. Effect of phase transition on the surface potential of the BaTiO

 (100) surface by variable temperature scanning surface potential microscopy. *Journal of Applied Physics* **2000**, *87* (8), 3950-3957. DOI: Doi 10.1063/1.372440.
(42) Liu, Y.; Roccapriore, K.; Checa, M.; Valleti, S. M.; Yang, J. C.; Jesse, S.; Vasudevan, R. K. AEcroscopy: a software–hardware framework empowering microscopy toward automated and autonomous experimentation. *Small Methods* **2024**, *8* (10), 2301740.
(43) Raghavan, A.; Pant, R.; Takeuchi, I.; Eliseev, E. A.; Checa, M.; Morozovska, A. N.; Ziatdinov, M.; Kalinin, S. V.; Liu, Y. Evolution of Ferroelectric Properties in Sm x Bi1–x FeO3 via Automated Piezoresponse Force Microscopy across combinatorial spread libraries. *ACS nano* **2024**, *18* (37), 25591-25600.
(44) Smith, B. R.; Pant, B.; Liu, Y.; Liu, Y.-C.; Yang, J.-C.; Jesse, S.; Khojandi, A.; Kalinin, S. V.; Cao, Y.; Vasudevan, R. K. Physics-informed models of domain wall dynamics as a route for autonomous domain wall design via reinforcement learning. *Digital Discovery* **2024**, *3* (3), 456-466.
(45) Liu, Y.; Kelley, K. P.; Vasudevan, R. K.; Zhu, W.; Hayden, J.; Maria, J. P.; Funakubo, H.; Ziatdinov, M. A.; Trolier-McKinstry, S.; Kalinin, S. V. Automated experiments of local non-linear behavior in ferroelectric materials. *Small* **2022**, *18* (48), 2204130.
(46) Liu, Y.; Yang, J.; Lawrie, B. J.; Kelley, K. P.; Ziatdinov, M.; Kalinin, S. V.; Ahmadi, M. Disentangling electronic transport and hysteresis at individual grain boundaries in hybrid perovskites via automated scanning probe microscopy. *ACS nano* **2023**, *17* (10), 9647-9657.
(47) Skidmore, C. H.; Spurling, R. J.; Hayden, J.; Baksa, S. M.; Behrendt, D.; Goodling, D.; Nordlander, J. L.; Suceava, A.; Casamento, J.; Akkopru-Akgun, B. Proximity ferroelectricity in wurtzite heterostructures. *Nature* **2025**, 1-6.

(48) Eliseev, E. A.; Morozovska, A. N.; Maria, J.-P.; Chen, L.-Q.; Gopalan, V. Thermodynamic Theory of Proximity Ferroelectricity. *Physical Review X* **2025**, *15* (2), 021058.
(49) Bi, F.; Bogorin, D. F.; Cen, C.; Bark, C. W.; Park, J.-W.; Eom, C.-B.; Levy, J. "Water-cycle" mechanism for writing and erasing nanostructures at the LaAlO3/SrTiO3 interface. *Applied Physics Letters* **2010**, *97* (17), Article. DOI: 10.1063/1.3506509.
(50) [https://www.wolfram.com/mathematica](https://www.wolfram.com/mathematica). (accessed 2025.

-